\documentclass[letterpaper, 10 pt, conference]{ieeeconf}  

\IEEEoverridecommandlockouts                              
\usepackage{cite}
\usepackage{amsmath,amssymb,amsfonts}
\usepackage{algorithm,algorithmic}
\usepackage{cases}
\usepackage{graphicx}
\usepackage{textcomp}
\usepackage{dsfont}
\usepackage{enumerate}

\usepackage{color}
\usepackage[caption=false,font=footnotesize]{subfig}

\newtheorem{lem}{Lemma}
\newtheorem{defin}{Definition}
\newtheorem{theorem}{Theorem}
\newtheorem{prop}{Proposition}

\renewcommand{\QED}{\QEDopen}

\newcommand{\iThetas}{{\it \Theta}^{\rm s}}
\newcommand{\iThetar}{{\it \Theta}^{\rm r}}
\newcommand{\ts}{\theta^{\rm s}}
\newcommand{\tr}{\theta^{\rm r}}
\newcommand{\mcal}[1]{\mathcal{#1}}
\newcommand{\sigs}{\sigma^{\rm s}}
\newcommand{\sigr}{\sigma^{\rm r}}
\newcommand{\sigss}{\sigma^{{\rm s}\ast}}
\newcommand{\sigrs}{\sigma^{{\rm r}\ast}}
\newcommand{\us}{u^{\rm s}}
\newcommand{\ur}{u^{\rm r}}
\newcommand{\ubs}{\bar{u}^{\rm s}}
\newcommand{\ubr}{\bar{u}^{\rm r}}
\newcommand{\Dus}{\Delta u^{\rm s}_{10}}
\newcommand{\Dur}{\Delta u^{\rm r}_{10}}
\newcommand{\pis}{\pi^{\rm s}}
\newcommand{\pir}{\pi^{\rm r}}
\newcommand{\gams}{\gamma^{\rm s}}
\newcommand{\gamr}{\gamma^{\rm r}}
\newcommand{\Dsigr}{\Delta \sigma^{\rm r}_{10}}

\DeclareMathOperator*{\argmax}{arg\,max}

\newcommand{\hs}{& \hspace{-3mm}}

\newcommand{\mcss}{\mathcal{S}^{\rm s}}

\newcommand{\mcsr}{\mathcal{S}^{\rm r}}

\title{\LARGE \bf
Epistemic Signaling Games\\ for Cyber Deception with Asymmetric Recognition
}

\author{Hampei Sasahara and Henrik Sandberg
\thanks{This work was supported by Swedish Research Council grant 2016-00861.}
\thanks{H. Sasahara and H. Sandberg are with Division of Decision and Control Systems, School of Electrical Engineering and Computer Science,
        KTH Royal Institute of Technology, Stockholm SE-100 44, Sweden
        {\tt\small \{hampei,hsan\}@kth.se}}%
}

\begin{document}

\maketitle
\thispagestyle{empty}
\pagestyle{empty}

\begin{abstract}

This study provides a model of cyber deception with asymmetric recognition represented by private beliefs.
Signaling games, which are often used in existing works, are built on the implicit premise that the receiver's belief is public information.
However, this assumption, which leads to symmetric recognition, is unrealistic in adversarial decision making.
For a precise evaluation of risks arising from cognitive gaps, this paper proposes epistemic signaling games based on the Mertens-Zamir model, which explicitly quantifies players' asymmetric recognition.
Equilibria of the games are analytically characterized with an interpretation.
\end{abstract}

\section{INTRODUCTION}

Cyber deception~\cite{Wang2018Cyber,Pawlick2019A}, which can obscure important data such as customer information or system architecture, is an emerging defense technology.
Examples of cyber deception include honeypots~\cite{The2004Know}, moving target defense~\cite{Zhuang2014Towards}, and obfuscation~\cite{Zhu2012Deceptive}.
Game theory offers mathematical models for strategic decision making~\cite{Farokhi2017Estimation,Saritas2017Quadratic,Miehling2019Strategic,Heydaribeni2019System,Nugraha2020Dynamic,Pirani2021Strategic}.
In particular, signaling games are often used for representing asymmetric players' knowledge, which arises especially in cyber deception~\cite{Carroll2011A,Ceker2016Deception,Pawlick2019Modeling}.
Signaling games are two-player games between a sender and a receiver, in which the sender's type is not known to the receiver.
At the beginning of the game, the receiver forms her prior belief on the sender's type.
Subsequently, the sender transmits her message, and the receiver updates her belief according to the message and chooses her action.
Using signaling games, reasonable actions of the attacker and the defender can mathematically be represented as equilibria.
The consequences of the game, such as the attack's impact and the deployed defense strategy's effectiveness, can be assessed in a quantitative manner by analyzing the equilibria.

An implicit assumption of traditional signaling games is that there exists a common prior, i.e., the receiver's prior belief is public information.
In the context of cyber deception, this assumption means that the defender exactly knows what the attacker believes.
Moreover, the attacker knows that the defender knows the attacker's belief.
This process repeats indefinitely, and their mutual beliefs are shared.
In this sense, the players' recognition is symmetric in traditional signaling games.
However, this assumption is obviously unrealistic.
As suggested by behavioral economics, recognition plays an important role in human's decision making~\cite{Camerar2004A}, and we may underestimate security risks without an adequate model that can describe \emph{asymmetric recognition.}

This study proposes \emph{epistemic signaling games} to resolve this issue.
The problem of asymmetric recognition has been pointed out in the general context of economics,
and the Mertens-Zamir model has been proposed to represent asymmetric recognition in epistemic game theory~\cite{Mertens1985Formulation,Dekel2015Epistemic,Zamir2009Bayesian}.
Using this model, we incorporate asymmetric recognition in signaling games.
We characterize equilibria of the proposed epistemic signaling games with an interpretation.

This paper is organized as follows.
In Sec.~\ref{sec:review}, we briefly review traditional signaling games.
Sec.~\ref{sec:epistemic} provides epistemic signaling games and analyze the equilibria.
Finally, Sec.~\ref{sec:conc} draws the conclusion.
The appendix contains the proofs.

\section{Brief Review: Traditional Signaling Games}
\label{sec:review}

\subsection{Example: Cyber Deception using a Honeypot}
\begin{figure}[t]
  \centering
  \includegraphics[width=0.98\linewidth]{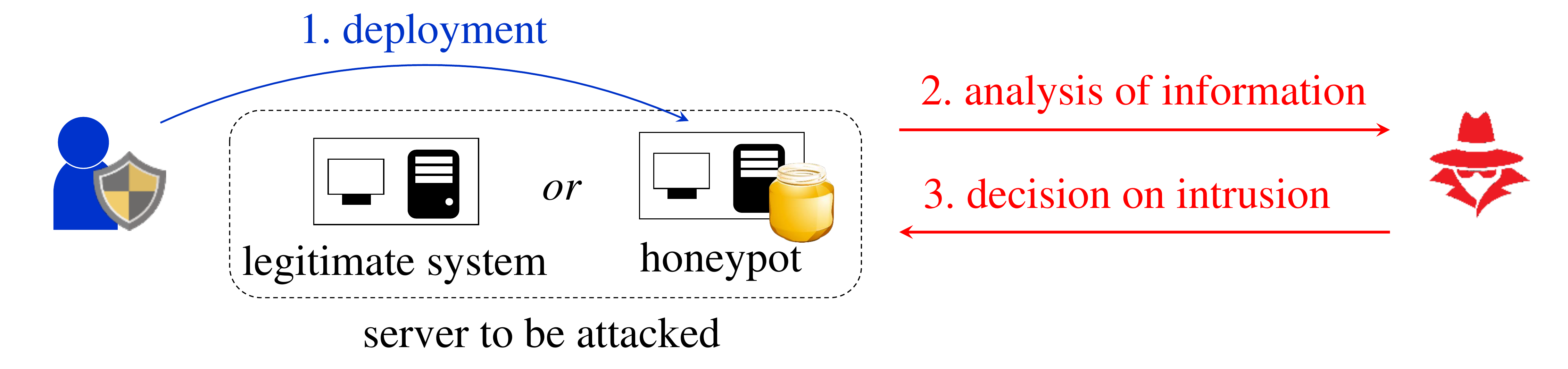}
  \caption{Example of cyber deception: the defender may deploy a honeypot, spending cost for its disguise, while the attacker decides whether to execute an intrusion by analyzing the available information.}
  \label{fig:ex_honey}
\end{figure}

An example is unauthorized access to a workstation that may be a honeypot, which is a system placed on a network to attract the attention of attackers.
A honeypot does not store any valuable data and collects information about the intruder's identity by alluring attackers~\cite{The2004Know}.
Once an attacker compromises a honeypot, the defender analyzes the adversarial actions in detail and utilizes the information to improve the network protection.
The adversarial decision making in the honeypot example is illustrated by Fig.~\ref{fig:ex_honey}.
The defender may deploy a honeypot, while the attacker may be able to identify the system by analyzing information caused by unusual behavior~\cite{Krawetz2004Anti,Fu2006On}.
In this example, the sender is a model of the defender who decides whether to spend cost for disguising a honeypot, while the receiver is a model of the attacker who decides whether to execute an intrusion by analyzing the information.

\subsection{Traditional Signaling Game Model}
\label{subsec:ex_model}

In signaling games, the sender's private information is referred to as its \emph{type}, which is denoted by $\ts \in \iThetas$.
For simplicity, we assume the type to be binary, i.e., $\iThetas=\{\ts_0,\ts_1\}$.
For example, $\ts_0$ and $\ts_1$ represent the legitimate system and the honeypot, respectively.
The true type is known to the sender but unknown to the receiver.

Given the type, the sender chooses a \emph{message} $m\in \mcal{M},$ which is assumed to be binary, i.e., $\mcal{M}=\{m_0,m_1\}$.
We refer to $m_i$ as an honest message when the type is $\ts_i$ and the other as a deceptive message for any $i\in\{0,1\}$.
We consider \emph{mixed strategies} and denote the sender's strategy by $\sigs \in \mcss$ such that $\sigs(m|\ts)$ gives the probability with which the sender sends the message $m$ when her type is $\ts$.
We refer to $m\in\mcal{M}$ as an \emph{on-path message} when $\sigs(m|\ts)>0$ for some $\ts\in\iThetas$, and as an \emph{off-path message} otherwise.
In the honeypot example, $m$ represents the service provided by the server.
The defender's choice of $m$ is a decision whether to disguise the system's behavior and to pay a cost, or not.
An example of the disguise is to replace a cheap honeypot that provides only an open service port with a sophisticated one that provides full functional support~\cite{Krawetz2004Anti}.

After receiving the message, the receiver chooses an \emph{action} $a\in\mcal{A}$, which is assumed to be binary, i.e., $\mcal{A}=\{a_0,a_1\}$.
In the honeypot example, $a_0$ and $a_1$ represent execution of the intrusion and withdrawal, respectively.
The receiver's mixed strategies are denoted by $\sigr \in \mcsr$.


Traditional signaling games assume existence of a \emph{common prior} on the type, i.e.,
the type is determined by nature according to a probability distribution over $\iThetas$, which is known to both players, at the beginning of the game.
The determined type is informed to the sender and the receiver updates her belief based on the prior distribution and the transmitted message.
The prior and posterior beliefs are denoted by $\pir(\ts)$ and $\pir(\ts|m),$ respectively.
In the next section we will revisit and discuss this assumption on existence of a common prior, which is standard in  Harsanyi's incomplete information games~\cite{Zamir2009Bayesian}.

Let $\us:\iThetas\times\mcal{M}\times\mcal{A}\to\mathbb{R}$ denote a utility function for the sender.
Similarly, let $\ur:\iThetas\times\mcal{A}\to\mathbb{R}$ denote a utility function for the receiver.
Note that the receiver's utility is independent of the message because the messaging cost is not owned by the receiver.
Throughout this paper, we assume that the utilities satisfy
\[
 \left\{\hspace{-2mm}
 \begin{array}{l}
 \us(\ts_0,m,a_0)<\us(\ts_0,m,a_1),\\
 \us(\ts_1,m,a_1)<\us(\ts_1,m,a_0),
 \end{array}
 \right.
 \left\{\hspace{-2mm}
 \begin{array}{l}
 \ur(\ts_0,a_0)>\ur(\ts_0,a_1),\\
 \ur(\ts_1,a_1)>\ur(\ts_1,a_0),
 \end{array}
 \right.
\]
for any $m\in\mcal{M}$.
This assumption means that, the attacker prefers $a_0$ if $\ts=\ts_0$ and $a_1$ if $\ts=\ts_1$, and the defender prefers the opposite if the message is fixed.
Moreover, we also assume that the sender's utility is symmetric with respect to her type for honest messaging, i.e., $\us (\ts_0,m_0,a_0) = \us (\ts_1,m_1,a_1)$ and $\us (\ts_0,m_0,a_1) = \us (\ts_1,m_1,a_0)$
to simplify the results.
Without this assumption, similar results are available by dividing the cases.

We suppose that deceptive messaging requires a cost.
The sender's utility function is assumed to be represented as
\[
\left\{
\begin{array}{l}
 \us(\ts_0,m_1,a)=\us(\ts_0,m_0,a)-C,\\
 \us(\ts_1,m_0,a)=\us(\ts_1,m_1,a)-C,
\end{array}
\right.\quad \forall a\in\mcal{A}
\]
where $C>0$ is the deceptive messaging cost.
In the honeypot example, $C$ may represent the cost for disguising the honeypot as a legitimate system.
This disguise can be achieved by replacing a cheap honeypot with a sophisticated, but expensive, one that can provide full functional support~\cite{Krawetz2004Anti}, for example.
Signaling games without deceptive messaging cost are referred to as cheap-talk signaling games~\cite{Carroll2011A}.
Our analysis can be extended to the cheap-talk case.

Reasonable strategies in signaling games are perfect Bayesian equilibria (PBE) where the players maximize their expected utilities and the receiver's belief is rationally updated according to Bayes' rule.
By investigating resulting PBE, as studied in~\cite{Carroll2011A,Ceker2016Deception}, we can assess security risks in a quantitative manner.
\section{Epistemic Signaling Games with Asymmetric Recognition}
\label{sec:epistemic}

\subsection{Implicit Assumption on Recognition in Traditional Model}

\begin{figure}[t]
\centering
\subfloat[][
 Symmetric recognition in the honeypot example:
 The attacker believes that the system of interest is legitimate with probability 0.8 and it is a honeypot with probability 0.2.
 The defender knows the attacker's belief, and the attacker knows that the defender knows the attacker's belief.
 This is the model used in the traditional signaling games.
 ]{\includegraphics[width=.98\linewidth]{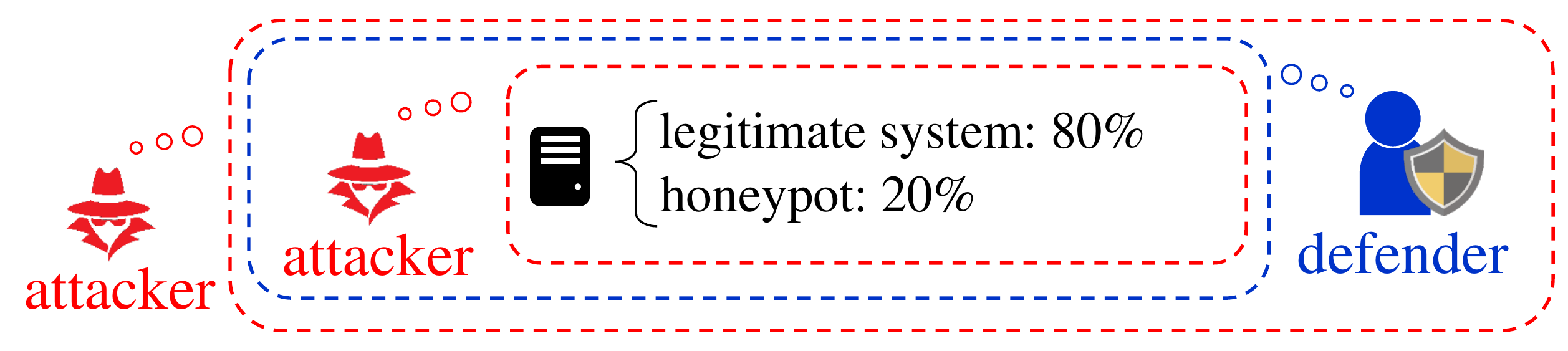}\label{fig:sym_reco}}
 \\
\subfloat[][
  Asymmetric recognition in the honeypot example:
  The attacker believes that the system of interest is legitimate with probability 0.2 and it is a honeypot with probability 0.8.
  On the other hand, the defender believes that the attacker believes that the system of interest is legitimate with probability 0.8 and it is a honeypot with probability 0.2 with probability 0.3 and that the attacker believes the system of interest is legitimate with probability 0.2 and it is a honeypot with probability 0.8 with probability 0.7.
  The attacker knows the defender's belief on the attacker's belief.
  This is the proposed model used in the epistemic signaling games.
]{\includegraphics[width=.98\linewidth]{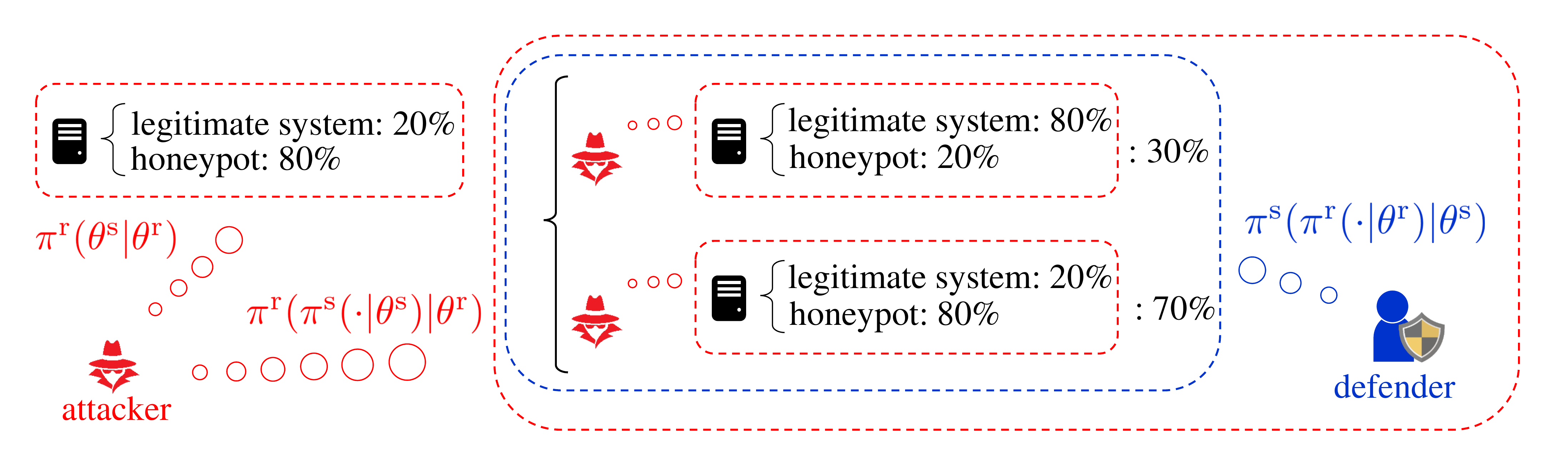}\label{fig:asym_reco}}
\caption{Symmetric and asymmetric recognitions in the honeypot example.}
\label{fig:recogs}
\end{figure}

An important feature of traditional signaling games is the existence of a common prior, under which the players precisely know what the opponent believes.
In other words, the existence of a common prior implicitly assumes \emph{symmetric recognition}.
However, in the security domain, the attacker's belief is not necessarily shared by the players, i.e., the attacker and the defender may possess \emph{asymmetric recognition}.
Figs.~\ref{fig:sym_reco} and~\ref{fig:asym_reco} illustrate symmetric and asymmetric recognitions in the honeypot example, respectively.

We briefly review the existing approach to resolving this issue used in the general context.
The most fundamental notion is \emph{belief hierarchy}, which has been introduced in epistemic game theory~\cite{Mertens1985Formulation,Dekel2015Epistemic,Zamir2009Bayesian}.
A belief hierarchy is formed as follows.
Let $\Delta(\mcal{X})$ denote the set of probability measures on $\mcal{X}$.
The first-order belief is given as $\pi_1 \in \Delta(\iThetas)$, which describes the attacker's belief on the system architecture.
The second-order belief is given as $\pi_2\in \Delta(\Delta(\iThetas))$, which describes the defender's belief on the attacker's first-order belief.
In a similar manner, the belief at any level is given, and the tuple of the beliefs at all levels is referred to as a belief hierarchy.

To handle incomplete information games without common prior and the resulting belief hierarchy, the \emph{Mertens-Zamir model} has been introduced~\cite{Mertens1985Formulation,Dekel2015Epistemic,Zamir2009Bayesian}.
The model considers \emph{type structure}, in which a belief hierarchy is embedded.
A type structure consists of players, sets of types, and beliefs.
In particular, for signaling games, a type structure can be given by
\begin{equation}\label{eq:type_str}
 \mcal{T}=(({\rm s},{\rm r}),(\iThetas,\iThetar),(\pis,\pir))
\end{equation}
where $({\rm s},{\rm r})$ represents the sender and the receiver,
$\iThetas$ and $\iThetar$ represent the sets of players' types,
and $\pis:\iThetar\times\iThetas\to[0,1]$ and $\pir:\iThetas\times\iThetar\to[0,1]$ represent the beliefs.
The value $\pis(\tr|\ts)$ denotes the sender's belief of the receiver's type $\tr$ when the sender's type is $\ts$, and $\pir(\ts|\tr)$ denotes the corresponding receiver's belief.
The first-order belief is given by $\pi_1(\ts)=\pir(\ts|\tr)$ for the true receiver's type $\tr\in\iThetar$, and the second-order belief is given by $\pi_2(\pir(\cdot|\tr)|\ts)=\pis(\tr|\ts)$ for the true sender's type $\ts\in\iThetas$ as long as there is a one-to-one correspondence between the receiver's types and her beliefs.
The higher-order beliefs are illustrated in Fig.~\ref{fig:asym_reco}.
In a similar manner, the belief at any level of the belief hierarchy can be derived from the type structure.
An important fact is that, for any reasonable belief hierarchy there exists a type structure that can generate the belief hierarchy of interest~\cite{Mertens1985Formulation,Dekel2015Epistemic,Zamir2009Bayesian}.
In this sense, the Mertens-Zamir model has the sufficient capability of describing any situation with asymmetric recognition.
For a formal discussion, see~\cite{Dekel2015Epistemic,Zamir2009Bayesian}.



\subsection{Epistemic Signaling Games with Asymmetric Recognition}

In this subsection, we propose \emph{epistemic signaling games} using the Mertens-Zamir model for describing adversarial decision making with asymmetric recognition.

In the honeypot example, the sender's type set is given by $\iThetas=\iThetas_{\rm sys}\times\iThetas_{\rm rec}$ where $\iThetas_{\rm sys}$ and $\iThetas_{\rm rec}$ represent the sets of the system's attribute, namely, a legitimate system or a honeypot, and of the defender's recognition, respectively.
For simplicity, we assume that $\iThetas_{\rm rec}$ is singleton, i.e., there is only one possible defender's recognition.
This implies that the defender's recognition is public information.
In this sense, this assumption corresponds to the worst case where the defender's belief is perfectly known to the attacker.
Under this assumption, \emph{the sender's belief is independent of her type}.
We denote the sender's belief by $\pis(\tr)$ instead of $\pis(\tr|\ts)$.
We also assume $\iThetas$ and $\iThetar$ in~\eqref{eq:type_str} to be binary, i.e., $\iThetas:=\{\ts_0,\ts_1\}$ and $\iThetar:=\{\tr_0,\tr_1\}$.

The contrasting ingredients of the traditional and our epistemic signaling games are listed in Table~\ref{table:ingredients}.
Accordingly, the sender's expected utility becomes
\[
 \begin{array}{l}
  \ubs(\sigs,\sigr|\ts)\\
  \displaystyle{
  :=
  \sum_{
  a\in\mcal{A}, m\in\mcal{M}, \tr\in\iThetar}
  \sigr(a|m,\tr)\pis(\tr)\sigs(m|\ts)\us(\ts,m,a)},
 \end{array}
\]
which depends on the sender's belief in contrast to the traditional one.
Similarly, the receiver's expected utility is given by
\[
 \ubr(\sigr|m,\tr):=\sum_{a\in\mcal{A},\ts\in\iThetas}\sigr(a|m,\tr)\pir(\ts|m,\tr)\ur(\ts,a),
\]
where $\pir(\ts|m,\tr)$ denotes the posterior belief.
For simplicity, we assume that the receivers with the type $\tr_0$ and $\tr_1$ prefer $a_0$ and $a_1$, respectively, when the sender's strategy is independent of her type, i.e., the message does not possess any information on sender's type.
In a mathematical form, we assume $\bar{u}^{\rm r}(a_0|\tr_0)>\bar{u}^{\rm r}(a_1|\tr_0)$ and $\bar{u}^{\rm r}(a_0|\tr_1)<\bar{u}^{\rm r}(a_1|\tr_1)$
where $\bar{u}^{\rm r}(a|\tr):=\sum_{\ts\in\iThetas}\pir(\ts|\tr)\ur(\ts,a)$.

\begin{table}[t]
\centering
\caption{Contrasting ingredients of epistemic signaling games}
\begin{tabular}{|c|c|c|} \hline
  & Traditional & Epistemic \\ \hline
 receiver's strategy & $\sigr(a|m)$ & $\sigr(a|m,\tr)$ \\ \hline
 sender's belief & none & $\pis(\tr|\ts)$ \\ \hline
 receiver's belief & $\pir(\ts)$ & $\pir(\ts|\tr)$ \\ \hline
\end{tabular}
\label{table:ingredients}
\end{table}

The solution concept is defined as follows.
\begin{defin}
A PBE of the epistemic signaling game is a strategy profile $(\sigss,\sigrs)$ and posterior belief $\pir(\ts|m,\tr)$ such that
\[
 \left\{
 \begin{array}{l}
 \sigss \in \argmax_{\sigs\in\mcss} \ubs(\sigs,\sigrs|\ts),\quad \forall \ts\in\iThetas,\\
 \sigrs \in \argmax_{\sigr\in\mcsr} \ubr(\sigr|m,\tr),\quad \forall m \in \mcal{M},\forall \tr\in\iThetar
 \end{array}
 \right.
\]
and
\[
 \pir(\ts|m,\tr) = \dfrac{\sigss(m|\ts)\pir(\ts|\tr)}{\sum_{\phi^{\rm s}\in\iThetas} \sigss(m|\phi^{\rm s})\pir(\phi^{\rm s}|\tr)}
\]
if $\sum_{\phi^{\rm s}\in\iThetas} \sigss(m|\phi^{\rm s})\pir(\phi^{\rm s}|\tr)\neq 0$.
\end{defin}

There are three categories of PBE: \emph{separating, pooling,} and \emph{partially-separating} PBE in traditional signaling games~\cite{Carroll2011A,Ceker2016Deception}.
At separating PBE, the senders having different types transmit opposite messages, i.e., $\sigs(m|\ts_0)=1-\sigs(m|\ts_1)\in\{0,1\}$ for any $m\in\mcal{M}$.
At pooling PBE, the sender's strategies are independent of the type, i.e., $\sigs(m|\ts_0)=\sigs(m|\ts_1)$ for any $m\in\mcal{M}$.
Otherwise, the PBE are referred to as partially-separating PBE.
The separating and pooling PBE describe two extreme cases.
In the honeypot example, the separating PBE mean that the legitimate system always provides full service while the honeypot always responds nothing.
In contrast, the pooling PBE mean that the system provides the same service regardless of its attribute, although the honeypot may record the malicious intrusion.

\subsection{Equilibrium Analysis of Epistemic Signaling Games}

For the equilibrium analysis, we introduce some notation.
Define
\[
 \left\{
\begin{array}{cl}
 \Dus(\ts_0) \hs := \us (\ts_0,m_0,a_1) - \us (\ts_0,m_0,a_0),\\
 \Dus(\ts_1) \hs := \us (\ts_1,m_1,a_1) - \us (\ts_1,m_1,a_0),
\end{array}
\right.
\]
which satisfy $\Dus(\ts_1)<0<\Dus(\ts_0)$.
Note that $\Dus(\ts_0)+\Dus(\ts_1)=0$ holds from the assumption on symmetry of the sender's utility.
We define the constant
$C':=C/\Dus(\ts_0),$
which is the cost normalized by $\Dus(\ts_0)$.
Moreover, define
$\Dur(\ts_0) := \ur(\ts_0,a_1)-\ur(\ts_0,a_0),$ and $\Dur(\ts_1) := \ur(\ts_1,a_1)-\ur(\ts_1,a_0),$
which satisfy $\Dur(\ts_0)<0<\Dur(\ts_1)$.

We first characterize best responses to a given opponent's strategy.
\begin{lem}\label{lem:BRs_costly}
Define $\gams:\mcsr \to\mathbb{R}$ and $\gamr:\mcss\times\mcal{M}\times\iThetar \to \mathbb{R}$ by
\begin{equation}\label{eq:gams_def}
 \begin{array}{rl}
 \gams(\sigr) \hs := \sum_{\tr\in\iThetar} \Dsigr(a_1|\tr)\pis(\tr),\\
 \gamr(\sigs,m,\tr) \hs := \sum_{\ts\in\iThetas} \sigs(m|\ts)\pir(\ts|\tr)\Dur(\ts),
 \end{array}
\end{equation}
where $\Dsigr(a|\tr):=\sigr(a|m_1,\tr)-\sigr(a|m_0,\tr).$
For a given receiver's strategy $\sigr$, the sender's best response $\sigss(m_0|\ts_0)$ is given by
\[
 \sigss(m_0|\ts_0) = \left\{
 \begin{array}{cl}
 1 & {\rm if}\ \gams(\sigr)<C',\\
 \alpha & {\rm if}\ \gams(\sigr) = C',\\
 0 & {\rm otherwise},
 \end{array}
 \right.
\]
for any $\alpha \in [0,1]$.
For $\ts_1$, the best response is given as the message opposite to the one by $\ts_0$ when $\gams(\sigr)\neq C'$ and otherwise given as an arbitrary number in $[0,1]$.
For a given sender's strategy $\sigs$, the receiver's best response $\sigrs(a_0|m,\tr)$ is given by
\[
 \sigrs(a_0|m,\tr) = \left\{
 \begin{array}{cl}
 1 & {\rm if}\ \gamr(\sigs,m,\tr) < 0,\\
 \alpha & {\rm if}\ \gamr(\sigs,m,\tr) = 0,\\
 0 & {\rm otherwise},
 \end{array}
 \right.
\]
when the message is on-path.
\end{lem}

The separating PBE and pooling PBE are given as follows.
\begin{theorem}\label{thm:sep_pool_PBE}
If $C\geq\Dus(\ts_0)$, all PBE are separating PBE given by
\begin{equation}\label{eq:sep_PBE}
 \left\{
 \begin{array}{lll}
 \sigss(m_0|\ts_0)\hs=\sigss(m_1|\ts_1)\hs=1,\\
 \sigrs(a_0|m_0,\tr)\hs=\sigrs(a_1|m_1,\tr)\hs=1,\quad \forall \tr\in\iThetar
 \end{array}
 \right.
\end{equation}
with the posterior belief $\pir(\ts_0|m_0,\tr)=\pir(\ts_1|m_1,\tr)=1$ for any $\tr\in\iThetar$.
If $C<\Dus(\ts_0)$ and $\pis(\tr_0)\geq C'$, there exist pooling PBE characterized by
\begin{equation}\label{eq:pool_PBE}
  \left\{
 \begin{array}{l}
 \sigss(m_0|\ts)=1,\quad \forall \ts\in\iThetas,\\
  \left\{
  \begin{array}{l}
  \sigrs(a_0|m_0,\tr_0)=1,\quad
  \sigrs(a_1|m_0,\tr_1)=1,\\
  \sigrs(a_1|m_1,\tr_0) \pis(\tr_0) - \sigrs(a_0|m_1,\tr_1) \pis(\tr_1) = C'
  \end{array}
  \right.
 \end{array}
 \right.
\end{equation}
with a suitable off-path posterior belief.
If $C<\Dus(\ts_0)$ and $\pis(\tr_1)\geq C'$, there exists PBE pooling at $m_1$, which can be characterized in a similar manner.
Furthermore, if $\pis(\ts_0)<C'$ and $\pis(\ts_1)<C'$, then the game admits no pooling PBE.
\end{theorem}

In Theorem~\ref{thm:sep_pool_PBE}, the claim on separating PBE implies that the sender is always honest when the cost $C$ is too high.
The other claim on pooling PBE implies that giving no information can become a reasonable sender's strategy when the cost $C$ is not too high.

Subsequently, we characterize partially-separating PBE.
As a preparation, we state the following lemma.
\begin{lem}\label{lem:gamr_prop}
The function $\gamr$ defined in~\eqref{eq:gams_def} satisfies
\begin{equation}\label{eq:gamr_prop1}
 \gamr(\sigs,m,\tr_0)<\gamr(\sigs,m,\tr_1),\quad \forall \sigs\in\mcss, \forall m\in\mcal{M}
\end{equation}
and
\begin{equation}\label{eq:gamr_prop2}
 \left\{
 \begin{array}{l}
 \gamr(\sigs,m_0,\tr_0)+\gamr(\sigs,m_1,\tr_0)< 0,\\
 \gamr(\sigs,m_0,\tr_1)+\gamr(\sigs,m_1,\tr_1)> 0,
 \end{array}
 \right. \forall \sigs \in \mcss.
\end{equation}
\end{lem}

Owing to Lemma~\ref{lem:gamr_prop}, we can reduce the number of possible cases.
\begin{lem}\label{lem:gamr_prop2}
The sender's optimal strategy $\sigss\in\mcss$ at any partially-separating PBE satisfies $\gamr(\sigss,m_0,\tr_0)<0<\gamr(\sigss,m_1,\tr_1).$
\end{lem}

\begin{figure}[t]
  \centering
  \includegraphics[width=0.98\linewidth]{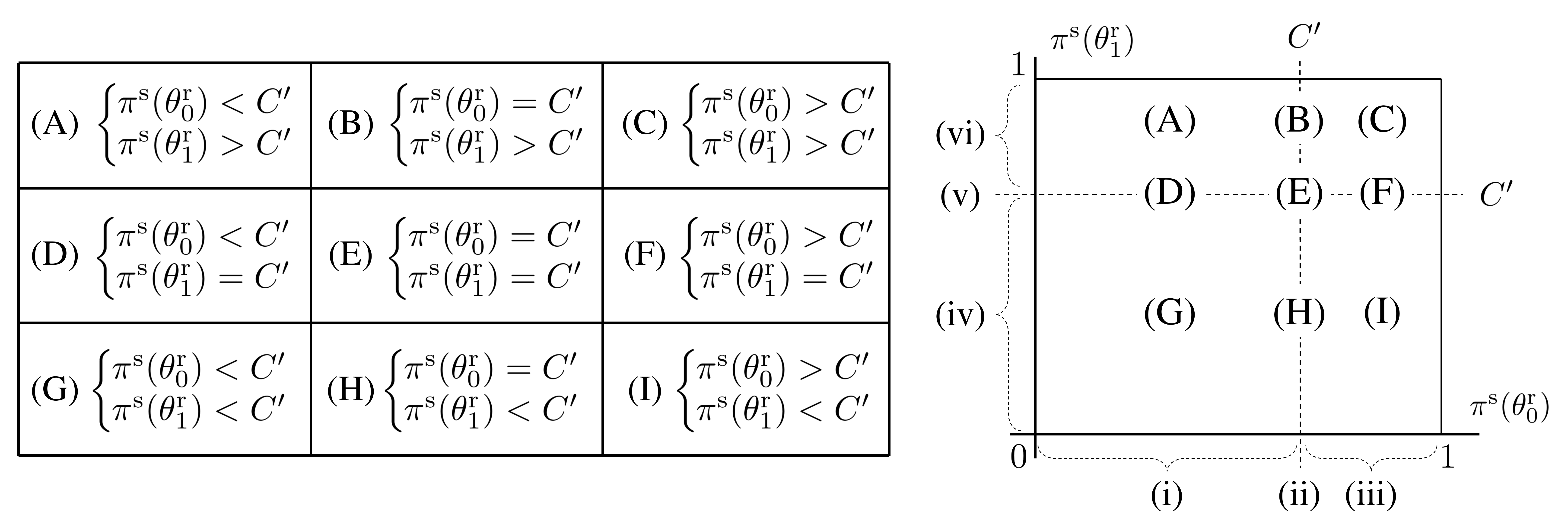}
  \caption{Regions of the sender's belief. The partially-separating PBE are determined from the region which the sender's belief belongs to.}
  \label{fig:regions}
\end{figure}

Using Lemma~\ref{lem:gamr_prop2}, we can characterize the partially-separating PBE.
\begin{theorem}\label{thm:par_sep_PBE}
Assume $C<\Dus(\ts_0)$.
In epistemic signaling games, there exist partially-separating PBE independent of the sender's belief characterized by
\begin{equation}\label{eq:PBE_independent}
 \left\{
 \begin{array}{l}
  \sigss: \left\{
   \begin{array}{l}
   \gamr(\sigss,m_0,\tr_0)<0,\quad
   \gamr(\sigss,m_1,\tr_0)=0,\\
   \gamr(\sigss,m_0,\tr_1)=0,\quad
   \gamr(\sigss,m_1,\tr_1)>0,
   \end{array}
  \right.\\
 \sigrs: \gams(\sigrs)=C',
 \end{array}
 \right.
\end{equation}
and the other partially-separating PBE dependent on the sender's belief are characterized by Table~\ref{table:PBE}, where the cases are specifically given in Fig.~\ref{fig:regions} and the equilibrium candidates are given in Table~\ref{table:PBE_cand}.
\end{theorem}


\begin{table}[t]
\centering
\caption{Partially-separating PBE of Epistemic Signaling Games}
\begin{tabular}{|l|l|l|} \hline
case (A): (i,vi)
&
case (B): (i,ii,iii,vi)
&
case (C): (iii,vi)
\rule[-2mm]{0mm}{5mm}
\\ \hline
case (D): (i,iv,v,vi)
&
case (E): (i,ii,iii,iv,v,vi)
&
case (F): (iii,iv,v,vi)
\rule[-2mm]{0mm}{5mm}
\\ \hline
case (G): (i,iv)
&
case (H): (i,ii,iii,iv)
&
case (I): (iii,iv)
\rule[-2mm]{0mm}{5mm}
\\ \hline
\end{tabular}
\label{table:PBE}
\end{table}

\begin{table*}[t]
\centering
\caption{Equilibrium Candidates}
\begin{tabular}{|cl|cl|} \hline 
  (i) &
 $
 \left\{
 \begin{array}{l}
 \sigss:
  \left\{
   \begin{array}{l}
   \gamr(\sigss,m_0,\tr_0)<0,\gamr(\sigss,m_1,\tr_0)>0,\\
   \gamr(\sigss,m_0,\tr_1)=0, \gamr(\sigss,m_1,\tr_1)>0,
   \end{array}
  \right.\\
  \sigrs:
  \left\{
   \begin{array}{l}
   \sigrs(a_0|m_0,\tr_0)=1, \sigrs(a_1|m_1,\tr_0)=1,\\
   \sigrs(a_1|m_0,\tr_1)=C'_1,\sigrs(a_1|m_1,\tr_1)=1
   \end{array}
  \right.
 \end{array}
 \right.
 $
 &
  (ii) &
 $
 \left\{
 \begin{array}{l}
 \sigss:
  \left\{
   \begin{array}{l}
   \gamr(\sigss,m_0,\tr_0)<0,\gamr(\sigss,m_1,\tr_0)>0,\\
   \gamr(\sigss,m,\tr_1)>0, \forall m\in\mathcal{M},
   \end{array}
  \right.\\
  \sigrs:
  \left\{
   \begin{array}{l}
   \sigrs(a_0|m_0,\tr_0)=1, \sigrs(a_1|m_1,\tr_0)=1,\\
   \sigrs(a_1|m,\tr_1)=1, \forall m\in\mathcal{M}
   \end{array}
  \right.
 \end{array}
 \right.
 $
 \rule[-7mm]{0mm}{16mm}
 \\ \hline
(iii) &
 $
 \left\{
 \begin{array}{l}
 \sigss:
  \left\{
   \begin{array}{l}
   \gamr(\sigss,m_0,\tr_0)<0,\gamr(\sigss,m_1,\tr_0)=0,\\
   \gamr(\sigss,m,\tr_1)>0, \forall m\in\mathcal{M},
   \end{array}
  \right.\\
  \sigrs:
  \left\{
   \begin{array}{l}
   \sigrs(a_0|m_0,\tr_0)=1, \sigrs(a_0|m_1,\tr_0)=C'_3,\\
   \sigrs(a_1|m,\tr_1)=1, \forall m\in\mathcal{M}
   \end{array}
  \right.
 \end{array}
 \right.
 $
 &
  (iv) &
 $
 \left\{
 \begin{array}{l}
 \sigss:
  \left\{
   \begin{array}{l}
   \gamr(\sigss,m_0,\tr_0)<0, \gamr(\sigss,m_1,\tr_0)=0,\\
   \gamr(\sigss,m_0,\tr_1)<0, \gamr(\sigss,m_1,\tr_1)>0,
   \end{array}
  \right.\\
  \sigrs:
  \left\{
   \begin{array}{l}
   \sigrs(a_0|m_0,\tr_0)=1, \sigrs(a_0|m_1,\tr_0)=C'_4,\\
   \sigrs(a_0|m_0,\theta^{\rm r}_1)=1, \sigrs(a_1|m_1,\tr_1)=1
   \end{array}
  \right.
 \end{array}
 \right.
 $
 \rule[-7mm]{0mm}{16mm}
 \\ \hline
 (v) &
 $
 \left\{
 \begin{array}{l}
 \sigss:
  \left\{
   \begin{array}{l}
   \gamr(\sigss,m,\tr_0)<0, \forall m\in\mathcal{M},\\
   \gamr(\sigss,m_0,\tr_1)<0, \gamr(\sigss,m_1,\tr_1)>0,
   \end{array}
  \right.\\
  \sigrs:
  \left\{
   \begin{array}{l}
   \sigrs(a_0|m,\tr_0)=1, \forall m\in\mathcal{M},\\
   \sigrs(a_0|m_0,\theta^{\rm r}_1)=1, \sigrs(a_1|m_1,\tr_1)=1
   \end{array}
  \right.
 \end{array}
 \right.
 $
 &
  (vi) &
 $
 \left\{
 \begin{array}{l}
  \sigss:
  \left\{
   \begin{array}{l}
   \gamr(\sigss,m,\tr_0)<0, \forall m\in\mathcal{M},\\
   \gamr(\sigss,m_0,\tr_1)=0, \gamr(\sigss,m_1,\tr_1)>0,
   \end{array}
  \right.\\
  \sigrs:
  \left\{
   \begin{array}{l}
   \sigrs(a_0|m,\tr_0)=1,\forall m\in\mathcal{M},\\
   \sigrs(a_1|m_0,\tr_1)=C'_6,\sigrs(a_1|m_1,\tr_1)=1
   \end{array}
  \right.
 \end{array}
 \right.
 $
 \rule[-7mm]{0mm}{16mm}
 \\ \hline
 \multicolumn{4}{|l|}{
 Constants: $C'_1:=(1-C')/\pis(\tr_1),C'_3:=1-C'/\pis(\tr_0),C'_4:=(1-C')/\pis(\tr_0),C'_6:=1-C'/\pis(\tr_1)$
 \rule[-2mm]{0mm}{5mm}
 }
 \\ \hline
\end{tabular}
\label{table:PBE_cand}
\end{table*}

\begin{figure}[t]
  \centering
  \includegraphics[width=0.98\linewidth]{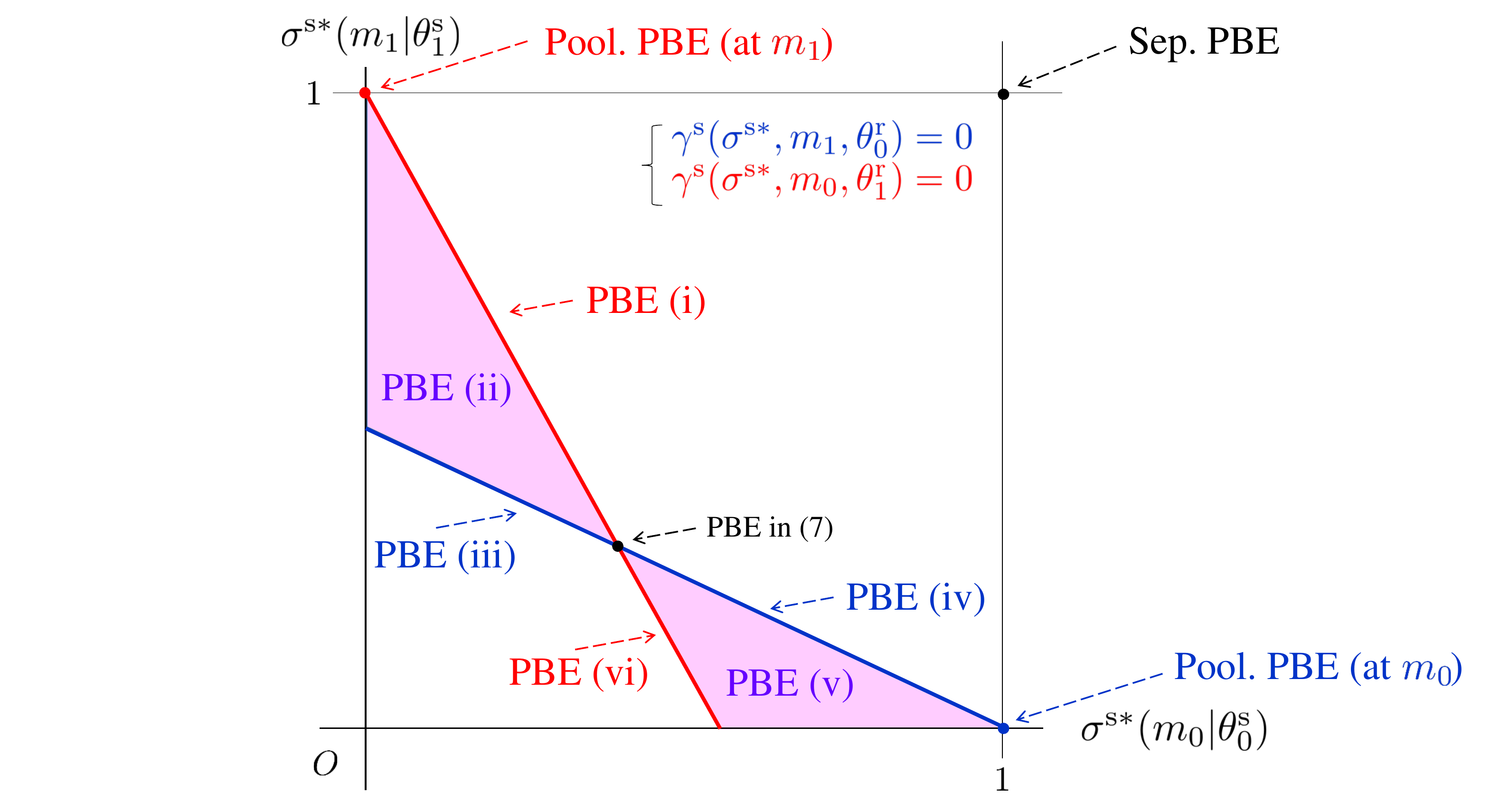}
  \caption{
  Possible equilibrium strategies of the sender, as given by Theorem~\ref{thm:par_sep_PBE}.
  The horizontal and vertical axes are the sender's mixed strategies of honest messaging for $\ts_0$ and $\ts_1$, respectively.
  The points $(1,1)$, $(1,0)$, and $(0,1)$ are the separating PBE, and the PBE pooling at $m_0$ and $m_1$, respectively.
  The red and blue lines depict the regions of sender's strategies that satisfy $\gams(\sigss,m_0,\tr_1)=0$ and $\gams(\sigss,m_1,\tr_0)=0$, respectively.
  The intersection is associated with the PBE in~\eqref{eq:PBE_independent}.
  The upper left segment of the red line is associated with the PBE (i), and the lower right is associated with the PBE (vi).
  The upper left segment of the blue line is associated with the PBE (iii), and the lower right is associated with the PBE (iv).
  }
  \label{fig:sender_strategy_single}
\end{figure}

An interpretation of the derived PBE can be given as follows.
First, the receiver's strategy is determined from the condition $\gams(\sigrs)=C'$ at any PBE, which means that the receiver always balances the sender's expected utilities corresponding to the messages $m_0$ and $m_1$ independently of her type.
On the other hand, the sender's equilibrium strategy regions are illustrated in Fig.~\ref{fig:sender_strategy_single} where the solid line segments depict the conditions $\gams(\sigss,m_1,\ts_0)=0$ and $\gams(\sigss,m_0,\ts_1)=0$.
The endpoints $(1,0)$ and $(0,1)$ correspond to the pooling PBE, and the intersection corresponds to the PBE independent of the sender's belief given by~\eqref{eq:PBE_independent}.
It can be observed that the region at the PBE (i) and (vi) are connected to that at the PBE pooling at $m_1$.
This PBE is taken when the sender believes that the receiver's type is $\tr_1$ with a high probability.
In this sense, the PBE (i) and (vi) are reasonable consequences when the sender believes $\tr_1$.
Similarly, the PBE (iii) and (iv) can be reasonable when the sender believes $\tr_0$.

\begin{figure*}[t]
  \centering
  \includegraphics[width=0.98\linewidth]{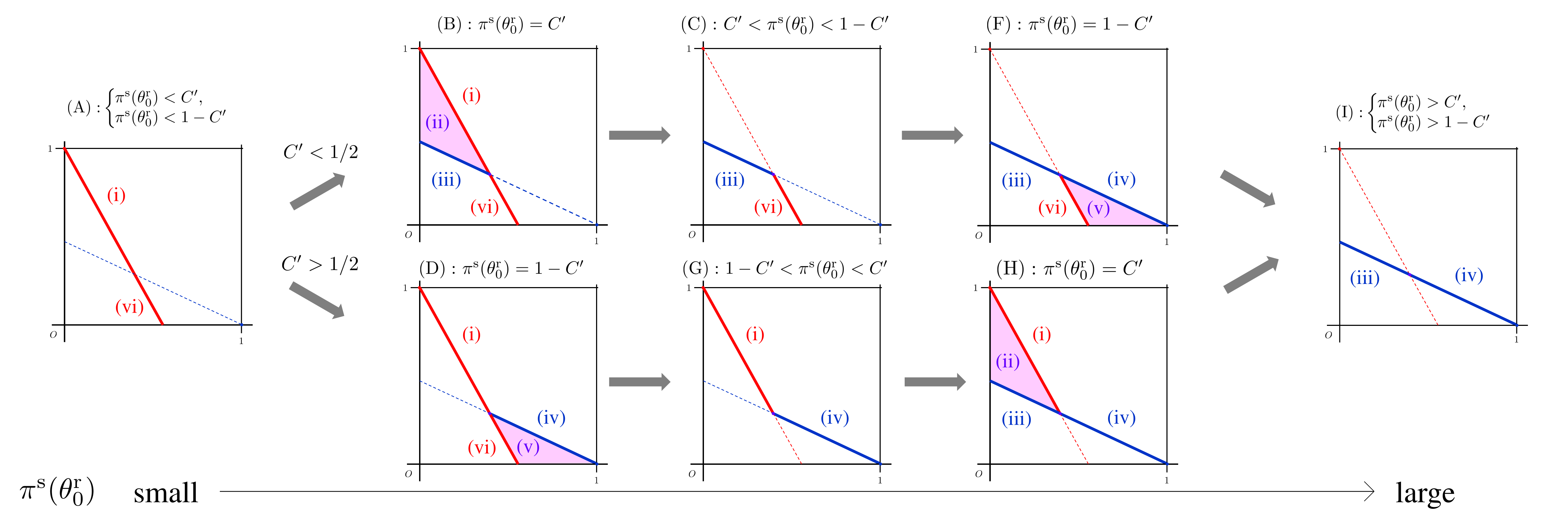}
  \caption{
  Transition of the sender's equilibrium strategies for increasing $\pis(\tr_0)$.
  The sender's strategies are given by the region depicted by the red solid line when $\pis(\tr_0)$ is small.
  As $\pis(\tr_0)$ increases, the region shifts to the one depicted by the blue solid line.
  }
  \label{fig:sender_strategy}
\end{figure*}

Fig.~\ref{fig:sender_strategy} illustrates the transition of the sender's equilibrium strategy regions for increasing $\pis(\tr_0)$.
Starting with $\pis(\tr_0)$ close to zero, the resulting PBE are (i) and (vi).
When $\pis(\tr_0)=C'$ and $C'<1/2$, the PBE (ii) and (iii) are additionally admitted.
When $\pis(\tr_0)$ slightly increases, the possible strategy region switches from the PBE (i) to (iii) through (ii).
It can be observed that, for such a moderate belief, the sender takes both $\tr_0$ and $\tr_1$ into account.
When $\pis(\tr_0)$ increases more, the region switches again and reaches the PBE (iii) and (iv).
A similar transition can be observed when $C'>1/2$.
Note that the separating PBE at the top right of Fig.~\ref{fig:sender_strategy_single} is the most honest strategy, and thus, the closer to the bottom left the strategy is, the more deceptive it is.
In this sense, the strategies at the PBE (iii) can be regarded as more deceptive than those at (iv), although both are reasonable when the sender believes $\tr_0$.
Indeed, the PBE (iii) is taken when the deception cost is low.
A similar interpretation is obtained for the PBE (i) and (vi).
Finally, the PBE (ii) and (v) ``bridge'' the other PBE.

\section{Conclusion}
\label{sec:conc}

This study proposes epistemic signaling games, a novel model of cyber deception with asymmetric recognition, based on the Mertens-Zamir model.
The equilibria are analytically characterized.


\appendix

\begin{proof}
\emph{Proof of Lemma~\ref{lem:BRs_costly}:}
With a slight abuse of notation, the difference between the sender's expected utilities using $m_0$ and $m_1$ is
\[
\begin{array}{l}
 \ubs(m_0,\sigr|\ts_0)-\ubs(m_1,\sigr|\ts_0)\\
 \ = -\sum_{a\in\mcal{A},\tr\in\iThetar}\pis(\tr)\us(\ts_0,m_0,a)\Dsigr(a|\tr)+C\\
 \ = -\gams(\sigr)\Dus(\ts_0)+C,
\end{array}
\]
which leads to the given best response for $\ts_0$.
Similarly, the same criterion can be obtained for $\ts_1$.

For the receiver, with an on-path message $m$, we have
\[
\begin{array}{l}
 \ubr(a_0|m,\tr)-\ubr(a_1|m,\tr)\\
 \ =\sum_{\ts\in\iThetas}\pir(\ts|m,\tr)\Dur(\ts)\\
 \ = \gamr(\sigs,m,\tr)/\sum_{\phi^{\rm s}\in\iThetas}\{\sigs(m|\phi^{\rm s})\pis(\phi^{\rm s}|\tr)\},
\end{array}
\]
which leads to the given best response.
\end{proof}


\begin{proof}
\emph{Proof of Theorem~\ref{thm:sep_pool_PBE}:}
Assume $C\geq\Dus(\ts_0)$, i.e., $C'\geq1$.
Since $\gams(\sigr)\leq 1$ for any $\sigr\in\mcsr$ and $\ts\in\iThetas$,
sending an honest message is always optimal.
Hence, the separating PBE is given by~\eqref{eq:sep_PBE}.

Next, assume $C<\Dus(\ts_0)$ and $\pis(\tr_0)\geq C'$.
Take the sender's strategy $\sigs(m_0|\ts)=1$ for any $\ts\in\iThetas$.
Then $\gamr(\sigs,m_0,\tr_0)<0$ and $\gamr(\sigs,m_0,\tr_1)>0$.
Hence the receiver's best response to $\sigs$ for $m_0$ is given by $\sigrs$ in~\eqref{eq:pool_PBE}.
Now $\sigs$ becomes the best response to $\sigrs$ if and only if $\gams(\sigrs)=C'$, i.e., the off-path strategy satisfies the equation in~\eqref{eq:pool_PBE}.
There exists an off-path strategy that satisfies the condition if and only if $\pis(\tr_0)\geq C'$.
Therefore, the claim holds.
For the case where $C<\Dus(\ts_0)$ and $\pis(\tr_1)\geq C'$, the pooling PBE can be derived in a similar manner.
Finally, if both of $\pis(\tr_0)$ and $\pis(\tr_1)$ are less than $C'$, then there exist no off-path strategies that satisfy the necessary equation in~\eqref{eq:pool_PBE}.
Hence there exist no pooling PBE.
\end{proof}


\begin{proof}
\emph{Proof of Lemma~\ref{lem:gamr_prop}:}
We have
$\Delta \gamr_{10}(\sigs,m) = \sum_{\ts\in\iThetas}\sigs(m|\ts)\Dur(\ts)\Delta \pi^{\rm r}_{10}(\ts)$
with $\Delta \gamr_{10}(\sigs,m):=\gamr(\sigs,m,\tr_1)-\gamr(\sigs,m,\tr_0)$ and $\Delta \pi^{\rm r}_{10}(\ts):= \pir(\ts|\tr_1)-\pir(\ts|\tr_0)$.
Since $\Delta \pi^{\rm r}_{10}(\ts_1)=-\Delta \pi^{\rm r}_{10}(\ts_0)$, we have
\[
 \begin{array}{l}
 \Delta \gamr_{10}(\sigs,m)\\
 \ =
 \{
  \underbrace{\sigs(m|\ts_1)\Dur(\ts_1)}_{>0}
  \underbrace{-\sigs(m|\ts_0)\Dur(\ts_0)}_{>0}
  \}
  \underbrace{\Delta \pi^{\rm r}_{10}(\ts_1)}_{>0},
 \end{array}
\]
which leads to~\eqref{eq:gamr_prop1}.
Further, $\gamr(\sigs,m_0,\tr)+\gamr(\sigs,m_1,\tr)=\bar{u}^{\rm r}(a_1|\tr)-\bar{u}^{\rm r}(a_0|\tr),$
which leads to~\eqref{eq:gamr_prop2}.
\end{proof}


\begin{proof}
\emph{Proof of Lemma~\ref{lem:gamr_prop2}:}
Since there exists $m\in\mcal{M}$ and $\ts\in\iThetas$ such that $\sigss(m|\ts)\in(0,1)$, we have $\gams(\sigrs)=C'$.
If $\gamr(\sigss,m_0,\tr_0)\geq 0$, then $\gamr(\sigss,m_1,\tr_0)<0$ from~\eqref{eq:gamr_prop2}.
Thus $\sigrs(a_1|m_1,\tr_0)=0$ and $\Delta\sigma^{{\rm r}\ast}_{10}(a_1|\tr_0)\leq 0$.
In addition, if $\gamr(\sigss,m_0,\tr_0)\geq 0$, then $\gamr(\sigss,m_0,\tr_1)>0$ from~\eqref{eq:gamr_prop1}.
Thus $\sigrs(a_1|m_0,\tr_1)=1$ and $\Delta\sigma^{{\rm r}\ast}_{10}(a_1|\tr_1)\leq 0$.
Therefore $\gams(\sigrs)\leq 0 \neq C'$, which leads to a contradiction.
The other claim can be proven in a similar manner.
\end{proof}


\begin{proof}
\emph{Proof of Theorem~\ref{thm:par_sep_PBE}:}
From Lemma~\ref{lem:gamr_prop2}, the possible combinations of the criterion for the best response are given by the nine cases: $\gamr(\sigss,m_1,\tr_0)\gtreqless 0,\gamr(\sigss,m_0,\tr_1)\gtreqless 0$
with $\gams(\sigrs)=C'$, $\gamr(\sigss,m_0,\tr_0)<0$, $\gamr(\sigss,m_1,\tr_1)>0$.

We first show that the two cases
\[
\left\{
\begin{array}{l}
\gamr(\sigss,m_1,\tr_0)<0\\
\gamr(\sigss,m_1,\tr_1)>0
\end{array}
\right.,\quad
\left\{
\begin{array}{l}
\gamr(\sigss,m_1,\tr_0)>0\\
\gamr(\sigss,m_1,\tr_1)<0
\end{array}
\right.
\]
do not happen.
Assume that the former one holds.
Then $\Delta\sigrs_{10}(a_1|\tr_0)=\Delta\sigrs_{10}(a_1|\tr_1)=0$, and hence $\gams(\sigrs)=0<C'$.
Similarly, the latter one implies that $\gams(\sigrs)=1>C'$.
Those conditions lead to contradictions.

For the other cases, the PBE and their existence conditions are derived by a routine calculations.
For example, for the PBE~\eqref{eq:PBE_independent}, it suffices to find $\sigss$ that satisfy the equations.
This can be done using a standard linear algebra.
\end{proof}

\bibliographystyle{IEEEtran}
\bibliography{sshrrefs}

\end{document}